\documentclass[11pt,letterpaper]{article}
\pdfoutput=1
\usepackage{soul}
\usepackage{jheppub}
\usepackage[utf8]{inputenc}
\usepackage[T1]{fontenc}
\usepackage{amsfonts}
\usepackage{amsmath}
\usepackage{graphicx,subfigure}
\usepackage{tensor} 
\usepackage{mathtools} 
\allowdisplaybreaks[4]
\usepackage{amssymb}
\usepackage{euscript}
\usepackage{color}
\usepackage{braket}

\newcommand{\rh}{r_{\text{h}}}
\newcommand{\MB}{M_{\text{B}}}

\title{ \boldmath Destroying the event horizon of a nonsingular rotating quantum-corrected black hole}

\author[]{Si-Jiang Yang$^a$$^b$,}
\author[]{Yu-Peng Zhang$^a$$^b$,}
\author[]{Shao-Wen Wei$^a$$^b$,}
\author[]{Yu-Xiao Liu$^a$$^b$$^c$\footnote{Corresponding author}}

\emailAdd{yangsj18@lzu.edu.cn, zyp@lzu.edu.cn, weishw@lzu.edu.cn, liuyx@lzu.edu.cn}

\affiliation[a]{Lanzhou Center for Theoretical Physics, Key Laboratory of Theoretical Physics of Gansu Province, School of Physical Science and Technology, Lanzhou University, Lanzhou 730000, China \vspace{0.1cm}}

\affiliation[b]{Institute of Theoretical Physics $\&$ Research Center of Gravitation, Lanzhou University, Lanzhou 730000, China \vspace{0.1cm}}

\affiliation[c]{Key Laboratory for Magnetism and Magnetic of the Ministry of Education, Lanzhou University, Lanzhou 730000, China}

\abstract{The destruction of the event horizon of a nonsingular black hole, which is not prevented by the weak cosmic censorship conjecture, might provide us the possibility to access quantum regime of gravity inside black hole. We investigate the possibility of overspinning a nonsingular rotating quantum-corrected black hole by a test particle and a scalar field in this paper, and analyse the effect of the quantum parameter on the destruction of the event horizon. For the test particle injection, both extremal and near-extremal black holes cannot be overspun due to the existence of the quantum parameter. And the larger the quantum parameter the harder the black hole to be overspun. It seems that the quantum parameter acts as a protector to prevent the black hole to be destroyed. However, for the test scalar field scattering, both extremal and near-extremal black holes can be destroyed. Due to the loop quantum gravity correction, the angular velocity of the extremal black hole shifts from that of the extremal Kerr black hole. This provides a small range of wave modes to destroy the event horizon of the quantum-corrected black hole.}


\begin{document}
\maketitle

\flushbottom


\section{Introduction}\label{sec:intro}

Gravitational collapse inevitably leads to spacetime singularity. This is the celebrated Hawking-Penrose singularity theorem~\cite{Penr65,HaPe70}. The presence of spacetime singularity indicates the failure of gravitational theory. To protect the predictability of gravitational theories, Penrose proposed the weak cosmic censorship conjecture, which states that naked singularities cannot be formed through regular initial data~\cite{Penr69}. This indicates that spacetime singularities are hidden behind black hole event horizons and can never be seen by distant observers. The conjecture has become one of the foundations of black hole physics
though a general proof is still beyond reach. The weak cosmic censorship conjecture preserves the predictability of classical gravitational theories. However, it also forbids us to probe the high curvature regions inside the event horizon where quantum properties cannot be neglected. Thus the destruction of the event horizon might provide us the possibility to access quantum regime of gravity inside black holes.

The weak cosmic censorship conjecture has been put to the test by many ways, such as numerical evolution of the collapse of dust cloud or other matter fields~\cite{Chri84,OrPi87,ShTe91,Lemo92,Chop93,HSSLW20,SoHW21}, extensive nonlinear numerical simulations of perturbed black holes or black rings~\cite{CoIP21,EpGS20,CrSa17,FKLT17,FiKT16,LePr10,HeHM04}, numerical evolving of collision and merge of two black holes in four and higher dimensions~\cite{AnFS20,AELL19,AELL19a,SCPBHY09}. One of the intriguing ways to check the conjecture is through gedanken experiment to destroy the event horizon.

The general strategy of such gedanken experiments is to throw test particles or fields into the extremal or near-extremal black hole and check whether the event horizon exists or not for the final composite object. Pioneering work of Wald showed that particles causing the destruction of the event horizon of an extremal Kerr-Newman black hole just not be captured by the black hole~\cite{Wald74}. This indicates that the event horizon of an extremal Kerr-Newman black hole cannot be destroyed by test particles.  While further investigations suggest that a near-extremal black hole might ``jump over'' the extremal limit and become a naked singularity~\cite{Hube99,Hod02}. Hubeny showed that a near-extremal charged black hole can be overcharged by test particles~\cite{Hube99}. By extending Hod's result, Jacobson and Sotiriou found that the event horizon of a near-extremal Kerr black hole can be destroyed~\cite{JaSo09}. Similar counterexamples are also found in some alternative theories of gravity~\cite{GhFS19,GhMS21}. But after including backreaction and finite size effects, those counterexamples seemed to be rescued~\cite{BaCK10,BaCK11,ZVPH13,CoBa15,Gwak17,LiWL19}. Recently, by taking into account the second-order perturbations that comes from the matter fields, Sorce and Wald showed that the Kerr-Newman black hole cannot be destroyed~\cite{SoWa17}. Subsequent systematic works further support the result that the event horizon cannot be destroyed by this kind of gedanken experiment~\cite{SaJi21,ZhJi20,QYWR20,WaJi20,JiGa20,ChLN19,QuTW22}.

Besides the injection of particles to destroy an event horizon, the scattering of fields is also used to test the weak cosmic censorship conjecture. The scattering of a scalar field provides intriguing features due to superradiance where the scalar field extracts energy from a charged or rotating black hole. Semiz's work indicates that a classical complex scalar field cannot destroy the event horizon of an extremal dyonic Kerr-Newman black hole~\cite{Semi11}. By dividing the scattering process into a series of infinitesimal time intervals, Gwak showed that extremal and near-extremal Kerr-(anti) de Sitter black holes cannot be overspun by a test scalar field~\cite{Gwak18}. Further investigations for a series of other black holes indicate that both extremal and near-extremal black holes cannot be destroyed by a test scalar field~\cite{Gwak21,YWCYW20,LGCM21,NaQV16,GoNa20,Gwak20}. However, quantum mechanically, a near-extremal charged black hole might absorb a dangerous quanta through tunneling process to become a Kerr-Newman naked singularity~\cite{MaSi07,Hod08,MRSDV09}.

There is in general a consensus that black hole singularities are windows onto physics beyond general relativity. Usually, it is believed that spacetime singularities are just classical results and can be circumvented by correction of quantum gravity~\cite{BDHR21}.
Loop quantum gravity is one of the main nonperturbative approach to quantum theory of gravity. It provides a possible description of the gravitational field in regimes in which quantum effects cannot be neglected, such as singularities in black hole and in cosmological situations. Over the past few years, a lot of black hole solutions in loop quantum gravity have been constructed~\cite{BoMM21,BoMMa21,LZWJJAW20,AsOS18}. The most intriguing feature of these black holes is that the singularity inside the black hole is replaced by a transition surface connecting to a white hole, and the spacetime is regular everywhere~\cite{FuZh22,LiFZ22}. Recently, using the revised Newman-Janis algorithm, a nonsingular rotating black hole solution in loop quantum gravity has been constructed. This black hole solution can be regarded as a quantum correction of a Kerr black hole characterized by a small quantum parameter, and captures universal features of an effective nonsingular rotating black hole in loop quantum gravity~\cite{BrCY21}.

The destruction of a black hole event horizon might provide us the possibility to explore the interior of a black hole. For a black holes with singularity inside its event horizon, the destruction of the event horizon is prohibited by the weak cosmic censorship conjecture. However, for a nonsingular black hole, there is no central singularity inside the event horizon and the whole spacetime is regular. Hence, the destruction of the event horizon of a nonsingular black hole is not prohibited by the weak cosmic censorship conjecture~\cite{LiCo13}, and the destruction of such black hole does not lead to the loss of predictability.
Motivated by recent research of gedanken experiments of destroying the event horizon with test particles and fields, we try to investigate the destruction of the event horizon of the nonsingular rotating black hole in loop quantum gravity and explore the effects of the quantum parameter on the destruction of the event horizon.

The outline of the paper is as follows. In Sec.~\ref{2}, we review the nonsingular rotating black hole in loop quantum gravity briefly. In Sec.~\ref{3} and Sec.~\ref{4}, we try to destroy the event horizon of the extremal and near-extremal nonsingular rotating black holes in loop quantum gravity by a test particle and a scalar field respectively, and discuss the effects of the quantum parameter on the destruction of the event horizon. The last section is devoted to discussion and conclusion.


\section{The nonsingular rotating black hole in loop quantum gravity}\label{2}

Starting at a static spherically symmetric black hole in loop quantum gravity as a seed metric, Brahma et al. constructed a rotating black hole in loop quantum gravity using the revised Newman-Janis algorithm. The metric captures universal features of an effective nonsingular rotating black hole in loop quantum gravity~\cite{BrCY21}. The metric for the black hole in Boyer-Lindquist coordinates can be written in the form~\cite{BrCY21}
\begin{equation}
  ds^2=-\left(1-\frac{2Mb}{\rho^2}\right)dt^2-\frac{4aMb\sin^2\theta}{\rho}dtd\phi
  +\rho^2d\theta^2
  +\frac{\rho^2}{\Delta}dr^2+\frac{\Sigma\sin^2\theta}{\rho^2}d\phi^2, \label{metric}
\end{equation}
where the metric functions are
\begin{align}
\notag \Delta & =8A_\lambda \MB^2\tilde{a}b^2+a^2,  &\quad \Sigma &=(b^2+a^2)^2-a^2\Delta\sin^2\theta,  \\
   M&=\frac{1}{2}b(1-8A_\lambda \MB^2\tilde{a}),   &\quad  \rho^2 & =b^2+a^2\cos^2\theta,
\end{align}
with
\begin{eqnarray}
   b^2(x) &=&\frac{A_\lambda}{\sqrt{1+x^2}} \frac{\MB^2(x+\sqrt{1+x^2})^6+\MB^2}{(x+\sqrt{1+x^2})^3}, \label{bx2}\\
   \tilde{a}(x)  & =&\left(1-\frac{1}{\sqrt{2A_\lambda}}\frac{1}{\sqrt{1+x^2}}\right)\frac{1+x^2}{b(x)^2},
\end{eqnarray}
and $x=r/(\sqrt{8A_\lambda}\MB) $, where $\MB $ corresponds to the Dirac observable in the model, and $A_\lambda=(\lambda_k/\MB^2)^{2/3}/2 $ is a nonnegative dimensionless parameter, where the quantum parameter $\lambda_k$ originates from holonomy modifications~\cite{BoMM21,BoMMa21} and it is directly related to the fundamental area gap in the theory~\cite{BrCY21}.

The metric describes a symmetric bounce spacetime with a radial variable $r\in (-\infty,+\infty)$, and $r=0$ is the transition surface. The metric approaches a Kerr spacetime asymptotically at $| r |\to +\infty$, and recovers to a static spherically symmetric black hole solution in loop quantum gravity when the spin parameter $a$ vanishes. Moreover, the metric describes a flat spacetime for $a=0$ and $M=0$ \cite{GSSW20}, satisfying an essential consistency check lacking in some quantum gravity inspired solutions~\cite{BrCY21,HoMP10}.

When the quantum parameter $A_\lambda$ vanishes, the metric describes a Kerr spacetime.
Different from a Kerr spacetime, the classical singular ring is replaced by a timelike transition surface induced from nonperturbative quantum correction and the spacetime is regular everywhere for nonvanishing quantum parameter $A_\lambda$.

\begin{figure}
  \centering
  \includegraphics[width=8cm]{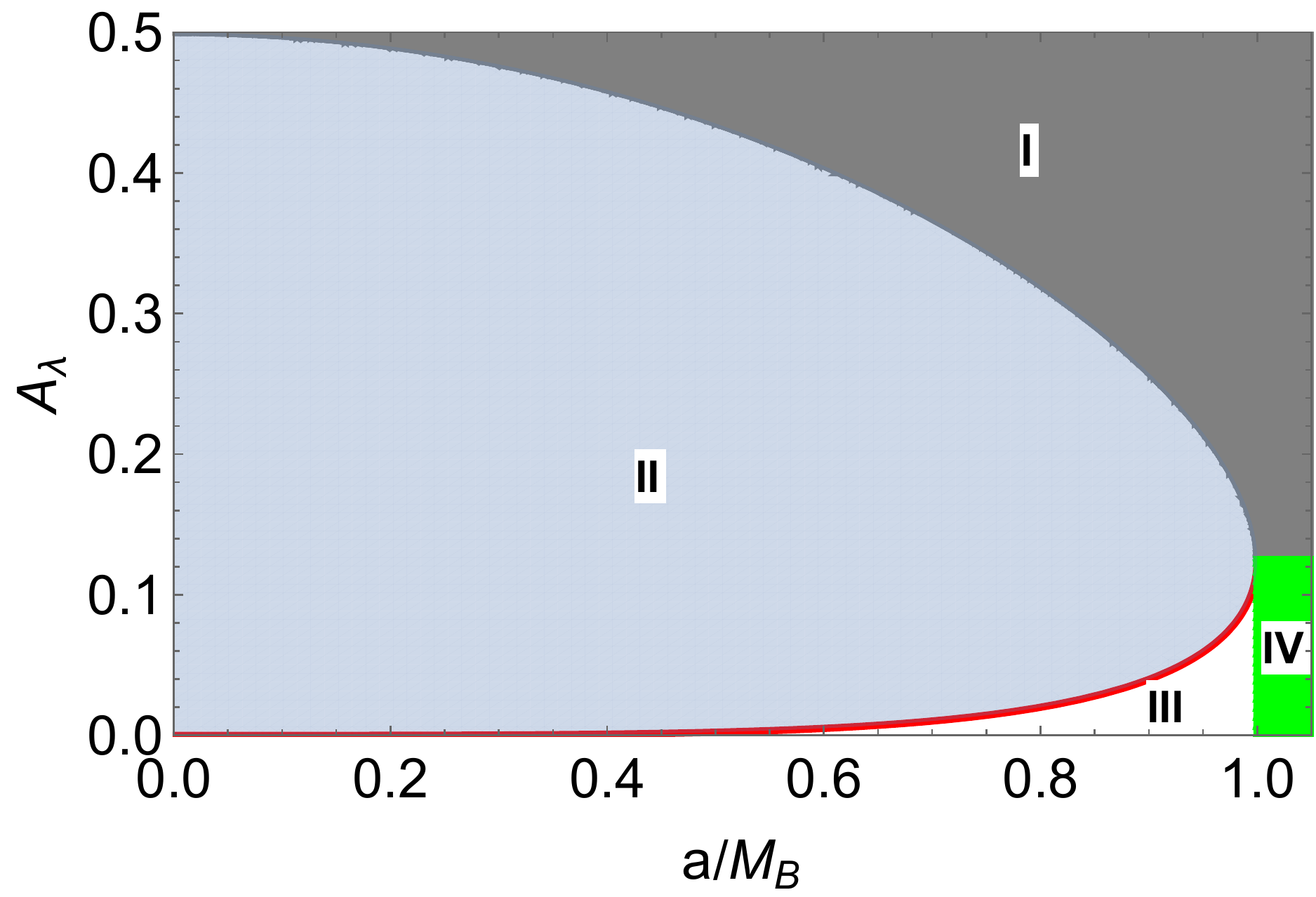}
  \caption{The dependence of the number of horizons on the parameters $A_\lambda$ and $a$.
 The numbers of horizons in regions \uppercase\expandafter{\romannumeral 1}, \uppercase\expandafter{\romannumeral2}, \uppercase\expandafter{\romannumeral3}, and \uppercase\expandafter{\romannumeral4} are 0, 1, 2 and 0, respectively.
  Region \uppercase\expandafter{\romannumeral 1} has been almost ruled out by the shadow size of M87* measured by EHT.
  \label{parameter}}
\end{figure}

The event horizon $\rh$ of the nonsingular rotating black hole in loop quantum gravity is defined by the equation $\Delta=0$, i.e.,
\begin{equation}\label{Horizon}
  \sqrt{8A_\lambda+\frac{\rh^2}{\MB^2}}=1\pm\sqrt{1-\frac{a^2}{\MB^2}},
\end{equation}
where the plus sign corresponds to the event horizon, while the minus sign to the inner horizon. Evidently, the number of the horizon strongly depends on the quantum parameter $A_\lambda$ and the spin parameter $a$. Figure~\ref{parameter} illustrates the dependence of the number of horizons on the parameters $A_\lambda$ and $a$.
From the figure, the parameter space can be divided into four regions. Region \uppercase\expandafter{\romannumeral1} describes a rotating wormhole without horizon and it has been almost ruled out by the shadow size of M87* measured by EHT~\cite{BrCY21}. For region \uppercase\expandafter{\romannumeral2}, the rotating loop  quantum gravity spacetime corresponds to a black hole with a horizon. While region \uppercase\expandafter{\romannumeral3} characterized by a small quantum parameter describes a rotating black hole with two horizons. From the observational implications of the shadow cast by this object, region \uppercase\expandafter{\romannumeral3} is the most physically relevant one for considering rotating black holes~\cite{BrCY21}. Region \uppercase\expandafter{\romannumeral4} is also characterized by a small quantum parameter but with large spin parameter $a/\MB>1$, hence there is no horizon. We focus on region \uppercase\expandafter{\romannumeral3} for considering the possibility of destruction of the event horizon, and we consider the spacetime with radial variable $r\geq0$.

The metric describes a rotating spacetime and it reduces to the Kerr metric for $r\to\infty$. Hence, the mass and angular momentum of the black hole are
\begin{align}\label{MandJ}
  M_{\text{ADM}} &= \lim_{r\to\infty} M=\MB, \\
  J& =\lim_{r\to\infty} Ma=\MB a.
\end{align}
The surface area of the black hole event horizon is
\begin{equation}\label{entropy}
  A=4\pi\left(b^2(\rh)+a^2\right),
\end{equation}
and the Hawking temperature of the event horizon is
\begin{equation}\label{Temp}
  T_{\text{H}}=\frac{\rh}{2\pi\left(a^2+b^2(\rh)\right)}\left(1-\frac{\MB}{\sqrt{8A_\lambda\MB^2+\rh^2}}\right).
\end{equation}
For an extremal nonsingular rotating black hole, the temperature is zero.
The metric describes a rotating black hole with angular velocity
\begin{equation}\label{angularV}
  \Omega_{\text{H}}=\frac{a}{a^2+b^2(\rh)}.
\end{equation}

\section{Overspinning the black hole with test particles}\label{3}

In this section, we explore the possibility of destroying the event horizon of the extremal or near-extremal nonsingular rotating black hole in loop quantum gravity by throwing a test particle into the black hole.

The horizon of the nonsingular rotating black hole is determined by Eq.~(\ref{Horizon}).
For $a\leq\MB$, the metric~(\ref{metric}) describes a black hole; while for $a>\MB$, it describes a rotating spacetime without event horizon.

To overspin the black hole, we only need to throw particles or fields with large angular momentum into the extremal or near-extremal black hole to make the final composite object with $J'>\MB'^2$.

A particle with mass $m$ moving in the nonsingular rotating loop quantum gravity spacetime is described by the geodesic equation

\begin{equation}\label{geodics}
  \frac{d^2x^\mu}{d\tau^2}+\Gamma^{\mu}_{\alpha \beta}\frac{dx^\alpha}{d\tau}\frac{dx^\beta}{d\tau}=0,
\end{equation}
which can be derived from the Lagrangian
\begin{equation}\label{Lagrangian}
\begin{split}
   L & =\frac{1}{2}mg_{\mu\nu}\frac{dx^\mu}{d\tau}\frac{dx^\nu}{d\tau} \\
     & =\frac{1}{2}mg_{\mu\nu}\dot{x}^\mu\dot{x}^\nu.
\end{split}
\end{equation}

We drop the particle from rest at infinity in the equator, then the particle will move in the equatorial plane. The energy $\delta E$ and angular momentum $\delta J$ of the particle are
\begin{subequations}\label{dEanddJ}
\begin{align}
  \delta E & =-P_t=-\frac{\partial L}{\partial \dot{t}} =-mg_{0\nu}\dot{x}^\nu, \label{dEanddJa}\\
  \delta J & =P_\phi=\frac{\partial L}{\partial \dot{\phi}} =mg_{3\nu}\dot{x}^\nu.\label{dEanddJb}
\end{align}
\end{subequations}

In the process of absorbing the particle, the changes of the mass and angular momentum of the black hole are
\begin{align}\label{ParameterChange}
  \MB\to \MB'&=\MB+\delta E,   \\
   J\to J'&=J+\delta J.
\end{align}

To investigate whether the event horizon of the black hole can be destroyed, we first find the condition for the particle with energy $\delta E$ and angular momentum $\delta J$ to enter the black hole, and then check whether such particles can destroy the event horizon thus expose its inner structure to outside observers.

The four-velocity of a massive particle is a timelike and unit vector
\begin{equation}\label{4Velocity}
  U^\mu U_\mu=g_{\mu\nu}\frac{dx^\mu}{d\tau}\frac{dx^\nu}{d\tau}=\frac{1}{m^2}g^{\mu\nu}P_\mu P_\nu=-1.
\end{equation}
Substituting the energy (\ref{dEanddJa}) and angular momentum (\ref{dEanddJb}) of the particle into the above equation, we get
\begin{equation}
  g^{00}\delta E^2-2g^{03}\delta J\delta E+g^{11}P_r^2+g^{33}\delta J^2=-m^2.
\end{equation}
Then the energy of the particle is
\begin{equation}
  \delta E=\frac{g^{03}}{g^{00}}\delta J-\frac{1}{g^{00}}\left[(g^{03})^2\delta J^2
  -g^{00}(g^{33}\delta J^2+g^{11}P_r^2+m^2)\right]^{\frac{1}{2}},\label{deltaE}
\end{equation}
where we have chosen the minus sign since the motion of the particle outside the event horizon should be timelike and future directed, which is
\begin{equation}\label{futureD}
  \frac{dt}{d\tau}>0,
\end{equation}
and it is equivalent to the requirement
\begin{equation}\label{Ee}
  \delta E>-\frac{g_{03}}{g_{33}}\delta J.
\end{equation}

If the particle enters the black hole, it must cross the event horizon. On the event horizon of the nonsingular rotating black hole, the condition becomes
\begin{equation}\label{UperB}
\begin{split}
    \delta J & < \frac{a^2+b^2(\rh)}{a}\delta E \\
     & =\frac{\delta E}{\Omega_{\text{H}}}.
\end{split}
\end{equation}
Intuitively, a particle with too large angular momentum just ``miss'' the black hole due to the centrifugal repulsion force. Thus, for the particle to be captured by the black hole, the angular momentum of the particle must satisfy
\begin{equation}\label{UperBound}
\delta J<\delta J_{\text{max}}=\frac{\delta E}{\Omega_{\text{H}}}.
\end{equation}

On the other hand, to overspin the black hole, we need
\begin{equation}\label{Lbound}
  J+\delta J>(\MB+\delta E)^2,
\end{equation}
which is
\begin{equation}\label{LboundJ}
  \delta J>\delta J_{\text{min}}=\delta E^2+2\MB \delta E+(\MB^2-J).
\end{equation}

If the two conditions (\ref{UperBound}) and (\ref{LboundJ}) are satisfied simultaneously, the event horizon of the black hole can be destroyed and the inner structure of the nonsingular rotating black hole in loop quantum gravity can be exposed to outside observers.

For an initial extremal nonsingular rotating black hole in loop quantum gravity, we have $\MB=a$. Then the event horizon of the black hole is
\begin{equation}\label{HExtrem}
  \rh=\sqrt{1-8A_\lambda}\MB.
\end{equation}

Since region \uppercase\expandafter{\romannumeral3} in Fig.~\ref{parameter} is the most physically relevant one for considering rotating black holes and this is the region we are interested in, the positive quantum parameter $A_\lambda$ satisfies
\begin{equation}\label{Alambda}
  8A_\lambda<1.
\end{equation}

To first order for the particle energy $\delta E$, only if the following two conditions are satisfied simultaneously
\begin{subequations}\label{2Condition}
  \begin{align}
  \delta J & \leq \delta J_{\text{max}}=\frac{\delta E}{\Omega_{\text{H}}},\label{2Conditiona}  \\
  \delta J & > \delta J_{\text{min}}=2\MB \delta E,\label{2Conditionb}
\end{align}
\end{subequations}
can the extremal nonsingular rotating black hole be overspun.

From Eqs.~(\ref{angularV}) and (\ref{bx2}), it is easy to calculate the angular velocity for the extremal nonsingular rotating black hole with $M_{\text{B}}=a$:
\begin{equation}\label{angularVelocity2}
  \frac{1}{\Omega_{\text{H}}}=2\MB(1-3A_\lambda),
\end{equation}
where $0\leq A_\lambda < 1/8$. Thus, it can be seen that
\begin{equation}\label{ExangularV}
  \frac{1}{\Omega_{\text{H}}}\leq2\MB.
\end{equation}
The equality is true only for vanishing quantum parameter $A_\lambda=0$, which corresponds to the extremal Kerr black hole.

Clearly, according to (\ref{ExangularV}), the two conditions  (\ref{2Conditiona}) and (\ref{2Conditionb}) cannot be satisfied simultaneously. Thus an extremal nonsingular rotating black hole in loop quantum gravity cannot be destroyed. Furthermore, we find that the larger the quantum parameter $A_\lambda$, the more difficult for the horizon to be destroyed. This means that the existence of the quantum parameter $A_\lambda$ makes the event horizon of the extremal nonsingular rotating black hole more difficult to be overspun by test particles.
Evidently, it is consistent with previous research that the centrifugal repulsion force is just great enough to prevent particles destroying the extremal black hole from being captured, and the quantum parameter $A_\lambda$ increases this tendency.

Adding the second order term $\delta E^2$ just increases the lower bound for the angular momentum, which makes the event horizon more difficult to be destroyed.

For an initial near-extremal nonsingular rotating black hole in loop quantum gravity, to first order in the energy of the particle $\delta E$, the conditions to overspin the black hole are
\begin{align}\label{NearE}
  \delta J & >\delta J_{\text{min}}=2\MB \delta E+(\MB^2-J), \\
  \delta J & <\delta J_{\text{max}}=\frac{1}{\Omega_{\text{H}}}\delta E=\frac{a^2+b^2(\rh)}{a} \delta E.
\end{align}

Define a small positive dimensionless parameter by
\begin{equation}\label{Depsilon}
  \frac{a^2}{\MB^2}=1-\epsilon^2.
\end{equation}
For a near-extremal nonsingular rotating loop quantum gravity black hole, we have $\epsilon\ll 1$.
From Eqs.~(\ref{angularV}), (\ref{Depsilon}), and (\ref{bx2}), we have
\begin{equation}
  \frac{1}{\Omega_{\text{H}}}-2\MB=\MB\left[-6A_{\lambda}+2\epsilon+(1-3A_{\lambda})\epsilon^2+{\cal O}(\epsilon^3) \right].
\end{equation}
Clearly, whether the near-extremal black hole can be destroyed by the test particle depends strongly on the quantum parameter $A_{\lambda}$. For $A_\lambda=0$, the metric describes a near-extremal Kerr black hole and it can be overspun as the previous research suggested~\cite{JaSo09}.
However, different from the previous research for the near-extremal nonsingular black holes in general relativity~\cite{FYTYL21}, the near-extremal nonsingular rotating black hole in loop quantum gravity cannot be overspun due to the positivity of the quantum parameter $A_\lambda$. And the larger the quantum parameter $A_\lambda$, the harder the black hole to be overspun. It seems that the quantum parameter $A_{\lambda}$ acts as a protector to prevent the destruction of the event horizon from test particles.

Hence, both the extremal and near-extremal nonsingular rotating black holes in loop quantum gravity cannot be overspun by test particles, and the larger the quantum parameter the harder the black hole to be destroyed.


\section{Overspinning the black hole with a test massive scalar field}\label{4}

Another method to destroy the event horizon is shooting a scalar field with large angular momentum into the extremal or near-extremal black hole. The gedanken experiment of scattering of a classical field into an extremal or near-extremal black hole to destroy the event horizon was proposed by Semiz~\cite{Semi11}. The research showed that a classical complex massive scalar field cannot destroy the extremal dyonic Kerr-Newman black hole. This method was further developed by others~\cite{DuSe13,SeDu15,Gwak18}.

In this section, we check the possibility of destroying the event horizon of the nonsingular rotating black hole in loop quantum gravity by shooting a massive classical scalar field into the extremal or near-extremal black hole and investigate the effect of the quantum parameter $A_\lambda$ on the destruction of the event horizon.


\subsection{The scattering for a massive scalar field}

We consider the scattering of a massive scalar field in the nonsingular rotating black hole. The massive scalar field $\Psi$ with mass $\mu$ minimally coupled to the gravity is governed by the Klein-Gordon equation
\begin{equation}\label{KGEq}
  \nabla_\mu\nabla^\mu\Psi-\mu^2\Psi=0,
\end{equation}
which can be written as
\begin{equation}
  \frac{1}{\sqrt{-g}}\partial_\mu\left(\sqrt{-g}g^{\mu\nu}\partial_\nu\Psi\right)-\mu^2\Psi=0.
\end{equation}

To solve the above Klein-Gordon equation, we make the following decomposition for the scalar field
\begin{equation}\label{Decomposition}
  \Psi(t,r,\theta,\phi)=e^{-i\omega t}R(r)S_{lm}(\theta)e^{im\phi},
\end{equation}
where $S_{lm}(\theta)$ is the spheroidal harmonic function, $m$ is the azimuthal harmonic index and it is a positive integer. Inserting the decomposition for the scalar field (\ref{Decomposition}) into the Klein-Gordon equation, we get the angular part of the equation
\begin{eqnarray}
   \frac{1}{\sin\theta}\frac{d}{d\theta}
   \bigg[\sin\theta\frac{dS_{lm}(\theta)}{d\theta}\bigg]
   -\bigg[a^2\omega^2\sin^2\theta  
   + \frac{m^2}{\sin^2\theta}+\mu^2a^2\cos^2\theta-\lambda_{lm}\bigg] S_{lm} =0, \label{angularpart}
\end{eqnarray}
and the radial equation
\begin{eqnarray}
  \frac{d}{dr}\left(\Delta\frac{dR}{dr}\right)+\left[\frac{(b^2+a^2)^2}{\Delta}\omega^2-\frac{4aMb}{\Delta}m\omega+\right. \left.\frac{m^2a^2}{\Delta}-\mu^2b^2-\lambda_{lm}\right]R(r)=0.\label{RadialE}
\end{eqnarray}

The solution to the angular equation is the spheroidal harmonic function~\cite{Seid89}. The separation constant $\lambda_{lm}$ is the eigenvalue of the spheroidal harmonic function and it is determined by the angular part of the Klein-Gordon equation. Here, we are more concerned with the radial equation, since the angular equation can be normalized to unity when we consider the energy and angular momentum fluxes scattering into the black hole.

To simplify the radial equation, we introduce the tortoise coordinate $r_*$ by
\begin{equation}\label{tortoise}
  \frac{dr}{dr_*}=\frac{\Delta}{b^2+a^2},
\end{equation}
where the tortoise coordinate $r_*$ ranges from $-\infty$ to $+\infty$ when the radial coordinate $r$ varies from the horizon $\rh$ to infinity. Thus it covers the whole space outside the event horizon.

Then the radial equation can be simplified to the following form
\begin{eqnarray}
  \frac{d^2R}{dr_*^2}+\frac{\Delta}{(b^2+a^2)^2}\frac{db^2}{dr}\frac{dR}{dr_*}
  + \left[\left(\omega-m\frac{a}{b^2+a^2}\right)^2
  -\frac{\mu^2b^2+\lambda_{lm}}{(b^2+a^2)^2}\Delta\right] R=0.
\end{eqnarray}

Near the event horizon, the above equation can be approximated as
\begin{equation}\label{RnearH}
  \frac{d^2R}{dr_*^2}+\left(\omega-m\frac{a}{b^2(\rh)+a^2}\right)^2R=0.
\end{equation}
Since the angular velocity of the black hole is
\begin{equation}
  \Omega_{\text{H}}=\frac{a}{b^2(\rh)+a^2},
\end{equation}
the radial equation~(\ref{RnearH}) can be rewritten as
\begin{equation}
  \frac{d^2R}{dr^2_*}+\left(\omega-m\Omega_{\text{H}}\right)^2R=0.
\end{equation}
The solution of the above equation is
\begin{equation}\label{Rsolution}
  R(r)\sim\exp\left[\pm i\left(\omega-m\Omega_{\text{H}}\right)r_*\right].
\end{equation}
We choose the negative sign, which corresponds to ingoing waves and it is the physically reasonable solution for black hole scattering.
Thus, the radial solution is
\begin{equation}
  R(r)=\exp\left[- i\left(\omega-m\Omega_{\text{H}}\right)r_*\right].
\end{equation}
Hence, the scalar field near the event horizon is
\begin{equation}\label{Fsolution}
  \Psi=\exp\left[- i\left(\omega-m\Omega_{\text{H}}\right)r_*\right]S_{lm}(\theta)e^{im\phi}e^{-i\omega t}.
\end{equation}

Having the solution for the scalar field near the event horizon, we can calculate the changes of the black hole parameters through the flux of the energy momentum for the scalar field.

We shoot a monotonic scalar field with mode $(l,m)$ into the black hole. The changes of the black hole parameters can be estimated from the energy flux and angular momentum flux of the scalar field during the scattering.

The energy momentum tensor of the scalar field is given by
\begin{equation}
  T_{\mu\nu}=\partial_{(\mu}\Psi\partial_{\nu)}\Psi^*-\frac{1}{2}g_{\mu\nu}\left(\partial_\alpha\Psi\partial^\alpha\Psi^*
  +\mu^2\Psi^*\Psi\right).
\end{equation}
The energy flux through the event horizon is
\begin{equation}\label{dEtodt}
\begin{split}
  \frac{dE}{dt} &= \int_{\text{H}}T^r_t\sqrt{-g}d\theta d\phi \\
     &= \omega(\omega-m\Omega_{\text{H}})\left[b^2(\rh)+a^2\right],
\end{split}
\end{equation}
and the angular momentum flux through the event horizon is
\begin{equation}\label{dJtodt}
\begin{split}
  \frac{dJ}{dt} &= \int_{\text{H}}T^r_\phi\sqrt{-g}d\theta d\phi \\
     &= m(\omega-m\Omega_{\text{H}})\left[b^2(\rh)+a^2\right],
\end{split}
\end{equation}
where we have used the normalization condition of the angular functions $S_{lm}(\theta)$ in the integration. From Eqs.~(\ref{dEtodt}) and (\ref{dJtodt}), for wave modes $\omega<m\Omega_{\text{H}}$, the energy and angular momentum fluxes are negative, which means that the scalar field extracts energy and angular momentum out of the black hole. This is called black hole superradiance~\cite{BrCP15}.

Then during a small time interval $dt$, the changes in the mass and angular momentum of the nonsingular rotating black hole in loop quantum gravity are
\begin{subequations}\label{dEdJ}
  \begin{align}
  dE & = \omega(\omega-m\Omega_{\text{H}})\left[b^2(\rh)+a^2\right]dt, \label{dEdJa} \\
  dJ & = m(\omega-m\Omega_{\text{H}})\left[b^2(\rh)+a^2\right]dt.\label{dEdJb}
\end{align}
\end{subequations}

Having the changes of the mass and angular momentum of the black hole during the scattering process, we can check whether the event horizon of the extremal or near extremal nonsingular rotating black hole can be destroyed.

\subsection{Overspinning the black hole with scalar field}

In this subsection, we try to overspin the black hole by shooting a monotonic classical test scalar field with frequency $\omega$ and azimuthal harmonic index $m$ into the extremal or near extremal nonsingular rotating black holes in loop quantum gravity, and investigate the effect of the quantum parameter $A_\lambda$ on the destruction of the event horizon.

Without loss of generality, we consider a small time interval $dt$. For a long time scattering process, we can divide it into a series of small time intervals and investigate each time interval individually only by changing the black hole parameters.

In the scattering process, an extremal or near-extremal black hole with mass $\MB$ and angular momentum $J$ absorbing a test scalar field with energy $dE$ and angular momentum $dJ$ becomes a composite object with mass $\MB'$ and angular momentum $J'$. To check whether the black hole is destroyed and hence expose its inner structure to outside observers, we only need to check the sign of $\MB'^2-J'$. If it is negative, there is no event horizon and the black hole is destroyed. Otherwise, the composite object is still a black hole.

After the scattering, we have
\begin{equation}
\begin{split}
   \MB'^2-J' & =\left(\MB+dE\right)^2-\left(J+dJ\right) \\
     & =\left(\MB^2-J\right)+2\MB dE-dJ.
\end{split}
\end{equation}
Inserting Eq.~(\ref{dEdJ}) into the above equation, we get
\begin{equation}
   \MB'^2-J' =\left(\MB^2-J\right)+
   2m^2\MB  \left[b^2(\rh)+a^2\right]
   \Big(\frac{\omega}{m}-
   \frac{1}{2\MB}\Big)
   \Big(\frac{\omega}{m}-\Omega_{\text{H}}\Big)dt.\label{FState}
\end{equation}
For an initial extremal nonsingular rotating black hole with
 $ \MB^2=J$, the above result (\ref{FState}) becomes
 \begin{equation}
    \MB'^2-J' =
   2m^2\MB \left(\frac{\omega}{m}-
   \frac{1}{2\MB}\right)
   \left(\frac{\omega}{m}-\Omega_{\text{H}}\right)\left[b^2(\rh)+a^2\right]dt.
 \end{equation}

As Eq.~(\ref{angularVelocity2}) shows, the angular velocity of the extremal nonsingular rotating  black hole can be written as
\begin{equation}
  \Omega_{\text{H}}=\frac{a}{a^2+b^2(\rh)}=\frac{1}{2\MB(1-3A_\lambda)}\geq\frac{1}{2\MB}.
\end{equation}
The equality holds only for vanishing quantum parameter, which corresponds to the angular velocity of an extremal Kerr black hole. Due to the loop quantum gravity correction, the angular velocity shifts from that of an extremal Kerr black hole. This has profound implications on the scattering of a scalar field for the nonsingular rotating black hole in loop quantum gravity. In the following, we show that the destruction of the event horizon for an extremal nonsingular rotating black hole is possible due to the angular velocity shifting from that of the Kerr black hole.

We shoot a scalar field with the following wave mode into the extremal black hole
\begin{equation}
  \frac{\omega}{m}=\frac{1}{2}\left(\frac{1}{2\MB}+\Omega_{\text{H}}\right).
\end{equation}
After absorbing the scalar field, the composite object has the following state
\begin{eqnarray}
  \MB'^2-J' = -\frac{1}{2}m^2\MB  
   \left(\Omega_{\text{H}}-\frac{1}{2\MB}\right)^2
  \left[a^2+b^2(\rh)\right]dt\leq0.
\end{eqnarray}
It equals to zero only for vanishing quantum parameter $A_\lambda$, and it shows that an extremal Kerr black hole absorbing the scalar field is still an extremal Kerr black hole. For positive quantum parameter $A_\lambda$, there is no event horizon after absorbing the scalar field and the black hole event horizon has been destroyed. After the horizon is destroyed by the scalar field, the overspinning black hole is in region \uppercase\expandafter{\romannumeral4} in Fig.~\ref{parameter}.  The result shows that an extremal nonsingular rotating black hole in loop quantum gravity can be overspun by a scalar field.

In fact, there is a small range of wave modes for the scalar field which can overspin the extremal nonsingular rotating black hole due to the nonzero of the quantum parameter $A_\lambda$. Clearly, for wave modes with
\begin{equation}\label{Wmode}
  \frac{1}{2\MB}<\frac{\omega}{m}<\frac{1}{2\MB (1-3A_\lambda)}=\Omega_{\text{H}},
\end{equation}
the final state of the composite object has
\begin{equation}
  \MB'^2-J'<0,
\end{equation}
which means that for wave modes satisfying Eq.~(\ref{Wmode}), an extremal nonsingular rotating black hole can be overspun.

From Eq.~(\ref{Wmode}), it is clear that the larger the quantum parameter $A_\lambda$, the wider the range of the wave modes can overspin the extremal black hole. The range of wave modes shrinks to zero for vanishing quantum parameter $A_\lambda$, which shows that an extremal Kerr black hole cannot be destroyed by a scalar field.

For a near-extremal nonsingular rotating black hole, we have
\begin{equation}
   \MB'^2-J' =\left(\MB^2-J\right)+
   2m^2\MB
   \left(\frac{\omega}{m}-
   \frac{1}{2\MB}\right)
   \left(\frac{\omega}{m}-\Omega_{\text{H}}\right)\left[b^2(\rh)+a^2\right]dt.\label{NEff}
\end{equation}
Still, we shoot a scalar field with wave mode with
\begin{equation}
  \frac{\omega}{m}=\frac{1}{2}\left(\Omega_{\text{H}}+\frac{1}{2\MB}\right)
\end{equation}
into the near-extremal nonsingular rotating black hole. Then Eq.~(\ref{NEff}) can be written in the form
\begin{equation}
   \MB'^2-J'  =\left(\MB^2-J\right)-\frac{1}{8\MB}m^2\Omega_{\text{H}}^2
      \left(\frac{1}{\Omega_{\text{H}}}-2\MB\right)^2
  \left[b^2(\rh)+a^2\right]dt. \label{NEf}
\end{equation}

We introduced a small positive dimensionless parameter in Eq.~\eqref{Depsilon} in Sec.~\ref{3} for the particle injection to destroy the black hole. Here, we also define a dimensionless parameter $\epsilon $ by
\begin{equation}
  \frac{a^2}{\MB^2}=1-\epsilon^2.
\end{equation}
Using power series expansion, we have
\begin{align}
  \MB^2-J & =\left[\frac{1}{2}\epsilon^2+{\cal O}(\epsilon^4)\right]\MB^2, \\
  \frac{1}{\Omega_{\text{H}}}-2\MB &=\left[-6A_\lambda+2\epsilon+{\cal O}(\epsilon^2)\right]\MB.
\end{align}
Thus, the final composite object has the state
\begin{equation}
   \MB'^2-J'=\left[\frac{1}{2}\epsilon^2+{\cal O}(\epsilon^4)\right]\MB^2-\frac{1}{8}m^2\MB \Omega_{\text{H}}^2
   \left[-6A_\lambda+2\epsilon+{\cal O}(\epsilon^2)\right]^2\left[b^2(\rh)+a^2\right]dt.
\end{equation}
Since the time interval $dt$ is of order $\epsilon$, for vanishing quantum parameter $A_\lambda$, the composite object has state $\MB^2-J'>0$, which means that a near-extremal Kerr black hole cannot be overspun by the test scalar field. The result is the same as the previous research~\cite{Gwak18}.

However, for nonzero quantum parameter $A_\lambda$, it is clear that
\begin{equation}
  \MB'^2-J'<0.
\end{equation}
This indicates that a near-extremal nonsingular rotating black hole in loop quantum gravity can be overspun by a classical test scalar field.

Hence, differing from extremal and near-extremal black holes in general relativity~\cite{Gwak18,YCWWL20,FYTYL21}, both extremal and near-extremal nonsingular rotating black holes in loop quantum gravity can be overspun by a test scalar field.

\section{Discussion and Conclusions}\label{7}

The weak cosmic censorship conjecture has become one of the foundations of black hole physics though a general proof of the conjecture has been out of reach yet. It might ultimately turn out to be true.
As the nonsingular rotating  black hole in loop quantum gravity has no central singularity, the destruction of the event horizon of this black hole does not lead to loss of the predictability of spacetime. Hence it does not violate the weak cosmic censorship conjecture. The destruction of the event horizon of a nonsingular black hole might provide us observable information about the interior of the black hole.

In this paper, we investigated the possibility of destroying the event horizon of a nonsingular rotating black hole in loop quantum gravity by a test particle and a scalar field, and analysed the effect of the quantum parameter $A_\lambda$ on the destruction of the black hole event horizon.

For the test particle injection, both extremal and near-extremal nonsingular rotating black holes  cannot be overspun. The larger the quantum parameter $A_\lambda$, the harder the black hole to be destroyed by a test particle. This differs from nonsingular black holes in general relativity, for which a near-extremal black hole can be destroyed by a test particle. It seems that the quantum parameter $A_\lambda$ acts as a protector to prevent a black hole to be destroyed by a test particle.

However, for the test scalar field scattering, both extremal and near-extremal nonsingular rotating  black holes can be destroyed. For an extremal black hole, the angular velocity shifts from that of the extremal Kerr black hole due to the loop quantum gravity correction. This provides a small range of wave modes to destroy the event horizon of the black hole in loop quantum gravity, and the range shrinks to zero for vanishing quantum parameter $A_\lambda$. The result shows that the quantum parameter makes the black hole event horizon more easy to be destroyed by a test scalar field.

\acknowledgments

 This work was supported in part by the National Natural Science Foundation of China (Grants No. 11875151, No. 12075103, No. 12047501, and No. 12105126), the China Postdoctoral Science Foundation (Grant No.2021M701531), the Fundamental Research Funds for the Central Universities (Grants No. lzujbky-2021-it34, No. lzujbky-2021-pd08), the 111 Project (Grant No. B20063), and Lanzhou City's scientific research funding subsidy to Lanzhou University.


\providecommand{\href}[2]{#2}\begingroup\raggedright\endgroup

\end{document}